# Extending the debate between Spearman and Wilson 1929: When do single variables optimally reproduce the common part of the observed covariances?

André Beauducel[1] & Norbert Hilger

University of Bonn, Germany

November 25, 2014

## Abstract

Because the covariances of observed variables reproduced from conventional factor score predictors are generally not the same as the covariances reproduced from the common factors, it is proposed to find a factor score predictor that optimally reproduces the common part of the observed covariances. It is shown that –under some conditions– the single observed variable with highest loading on a factor perfectly reproduces the non-diagonal observed covariances. This refers to Spearman's and Wilson's 1929 debate on the use of single variables as factor score predictors. The implications of this finding were investigated in a population based and in a sample based simulation study confirming that taking a single variable outperforms conventional factor score predictors in reproducing the observed covariances when the salient loading size and the number of salient loadings per factor are small. Implications of this finding for factor score predictors are discussed.

---

[1] Institute of Psychology, University of Bonn, Kaiser-Karl-Ring 9, 53111 Bonn, Germany, Email: beauducel@uni-bonn.de

## 1 Introduction

In 1929, Spearman proposed to add a single variable with perfect communality in order to avoid indeterminacy of the factor score. Wilson (1929) responded that then "…we may throw away all our scaffolding…" (p. 221) and that we should then identify the variable with the factor. However, weighting several variables in order to compute factor score predictors mirrors that factors are generally thought to represent some latent variable that affects several measured variables. There has been a considerable amount of debate on the meaning and usefulness of factor score predictors (Mulaik, 2010; Schönemann & Wang, 1972). A consequence of the indeterminacy of factor scores is that additional criteria have to be introduced in order to compute factor score predictors that might represent some aspects of the factor scores. Accordingly, several different factor score predictors representing different aspects of factor score predictors have been proposed (Grice, 2001a; Krijnen, 2006; McDonald & Burr, 1967). Starting from the observation that factor score predictors can be conceived as regression components (Schönemann & Steiger, 1976) the possibility to reproduce the common part of the observed covariances from the regression components corresponding to the factor score predictors is considered here. If factor score predictors represent the common factors well, they should approximately reproduce the same covariances as the common factors themselves. It is important to focus on the non-diagonal elements of the covariances, because the complete observed covariance matrix (including diagonal and non-diagonal elements) can best be reproduced by means of principal components analysis (Mulaik, 2010). However, an important advantage of common factors over principal components is that they reproduce the non-diagonal elements of the observed covariances. The focus on the reproduction of the non-diagonal elements of the covariance matrix is most obvious in the Minres-method (Comrey, 1962; Harman & Jones, 1966; Harman, 1976) of factor analysis, where the sum of squared non-diagonal residuals is minimized. Since factor score predictors should be as similar as possible to the common factors themselves, it is proposed here that the factor score predictors should reproduce the non-diagonal elements of the observed covariance matrix as well as the common factors do. This similarity of the reproduced observed covariance matrix might be regarded as a form of structural similarity, because it implies that the factor score predictors have the same quality as a hypothetical cause of the observed covariances as the common factors have. Thus, the criterion of structural similarity of the factor score predictor with the corresponding factor is added to the available criteria for the evaluation of factor score predictors. It should be noted that the criterion of structural similarity does not solve the problem of indeterminacy of factor score predictors. Searching for a factor score predictor that behaves like the factor in reproducing the observed covariances acknowledges that this factor score predictor needs not to be identical to the factor.

According to Grice (2001a) there are three criteria for the evaluation of factor score predictors: The correlation between the factor score predictor and the corresponding factor (validity), which should be at maximum. Moreover, the correlation of the factor score predictor with the non-corresponding factors should have the size of the correlation of the factor corresponding to the factor score predictor with the non-corresponding factors (univocality). Finally, the inter-correlation between the factor score predictors should be the same as the inter-correlation between the factors (correlational accuracy). These criteria are, of course, relevant for the evaluation of factor score predictors, but they are less closely tied to a central goal of common factor analysis, that is, that factors account for the covariances of observed variables (Preacher & MacCallum, 2003) and, more specifically that they account for the non-diagonal elements of the covariance matrix (Harman & Jones, 1966). Therefore, the criterion of structural similarity is proposed that allows to evaluate the factor score predictors as a basis for reproducing the (non-diagonal) covariances of the observed variables. This does, of course, not mean that the other criteria are irrelevant. Moreover, a factor score predictor that is perfectly correlated with the corresponding factor will reproduce the observed covariances exactly like the factor and will also have perfect univocality and correlational accuracy. However, when the correlation between the factor score predictor and the corresponding factor is not perfect, different factor score predictors might be optimal according to different criteria (Grice, 2001a).

Accordingly, the present paper investigates whether the single variable factor score predictor or conventional factor score predictors will best reproduce the non-diagonal part of the observed covariance matrix. This investigation is related to Spearman's (1929) idea to use a single variable with perfect communality as a factor score predictor, because such a variable will perfectly reproduce the non-diagonal observed covariances. Although Spearman proposed the use of single variables with perfect communality, it is explored to what degree and under which conditions single variables with a communality below one might adequately reproduce the non-diagonal observed covariances.

## 2 Definitions

The common factor model can be defined as

$$\mathbf{x} = \mathbf{\Lambda f} + \mathbf{e}, \tag{1}$$

where $\mathbf{x}$ is the random vector of observed variables of order $p$, $\mathbf{f}$ is the random vector of factor scores of order $q$, $\mathbf{e}$ is the unobservable random error vector of order $p$, and $\mathbf{\Lambda}$ is the factor pattern matrix of order $p$ by $q$. The common factor scores $\mathbf{f}$, and the error vectors $\mathbf{e}$ are assumed to have an expectation zero ($\epsilon[\mathbf{x}] = 0$, $\epsilon[\mathbf{f}] = 0$, $\epsilon[\mathbf{e}] = 0$). The variance of the factor scores is one, the covariance between the factors and the error scores is assumed to be zero ($\text{Cov}[\mathbf{f}, \mathbf{e}] = 0$). On this basis, the covariance matrix of observed variables $\mathbf{\Sigma}$ can be decomposed into

$$\mathbf{\Sigma} = \mathbf{\Lambda\Phi\Lambda}´ + \mathbf{\Psi^2}, \tag{2}$$

where $\mathbf{\Phi}$ represents the $q$ by $q$ factor correlation matrix and $\mathbf{\Psi^2}$ is a $p$ by $p$ diagonal matrix representing the covariance of the error scores $\mathbf{e}$ ($\text{Cov}[\mathbf{e}, \mathbf{e}] = \mathbf{\Psi^2}$). It is assumed that the diagonal of $\mathbf{\Psi^2}$ contains only positive values so that $\mathbf{\Psi^2}$ is nonsingular. Here, and in the following the model is considered for the population.

## 3 Results

Factor score predictors are based on the multiplication of a weight matrix $\mathbf{B}´$ with the observed variables $\mathbf{x}$. For example, the regression factor score predictor can be calculated as

$$\hat{\mathbf{f}}_\mathbf{r} = \mathbf{\Phi\Lambda}´\mathbf{\Sigma^{-1}x}, \tag{3}$$

with $\mathbf{B}=\mathbf{\Sigma^{-1}\Lambda\Phi}$. Accordingly, factor score predictors can be conceived as regression components in the sense of Schönemann and Steiger (1976). The regression component loading pattern $\mathbf{A}$ is

$$\mathbf{A} = \mathbf{\Sigma B(B}´\mathbf{\Sigma B)^{-1}}, \tag{4}$$

and the covariance between the regression components is $\mathbf{C} = \mathbf{B}'\boldsymbol{\Sigma}\mathbf{B}$. The covariance matrix of observed variables can be reproduced from the loading pattern $\mathbf{A}$ and the covariances $\mathbf{C}$ according to

$$\boldsymbol{\Sigma_r} = \mathbf{ACA}' = \boldsymbol{\Sigma}\mathbf{B}(\mathbf{B}'\boldsymbol{\Sigma}\mathbf{B})^{-1}\mathbf{B}'\boldsymbol{\Sigma}. \tag{5}$$

Beauducel (2007) showed that for Thurstone's regression factor score predictor (Thurstone, 1935), Anderson-Rubin factor score predictor (Anderson & Rubin, 1956), Bartlett's (1937) factor score predictor, and McDonald's (1981) factor score predictor, the reproduced covariance matrix is

$$\boldsymbol{\Sigma_r} = \boldsymbol{\Lambda}(\boldsymbol{\Lambda}'\boldsymbol{\Sigma}^{-1}\boldsymbol{\Lambda})^{-1}\boldsymbol{\Lambda}', \tag{6}$$

which is not identical to $\boldsymbol{\Lambda}\boldsymbol{\Phi}\boldsymbol{\Lambda}'$, the covariance matrix reproduced from the common factors. The difference between $\boldsymbol{\Sigma_r}$ and $\boldsymbol{\Lambda}\boldsymbol{\Phi}\boldsymbol{\Lambda}'$ becomes more obvious when Equation (6) is transformed according to Jöreskog (1969), Equation (10), so that

$$\boldsymbol{\Sigma_r} = \boldsymbol{\Lambda}(\boldsymbol{\Lambda}'\boldsymbol{\Psi}^{-2}\boldsymbol{\Lambda}\,(\mathbf{I} + \boldsymbol{\Phi}\boldsymbol{\Lambda}'\boldsymbol{\Psi}^{-2}\boldsymbol{\Lambda})^{-1})^{-1}\boldsymbol{\Lambda}'$$

$$= \boldsymbol{\Lambda}((\boldsymbol{\Lambda}'\boldsymbol{\Psi}^{-2}\boldsymbol{\Lambda})^{-1} + \boldsymbol{\Phi})\boldsymbol{\Lambda}'$$

$$= \boldsymbol{\Lambda}\boldsymbol{\Phi}\boldsymbol{\Lambda}' + \boldsymbol{\Lambda}(\boldsymbol{\Lambda}'\boldsymbol{\Psi}^{-2}\boldsymbol{\Lambda})^{-1}\boldsymbol{\Lambda}'. \tag{7}$$

Obviously, Equation (7) approaches $\boldsymbol{\Lambda}\boldsymbol{\Phi}\boldsymbol{\Lambda}'$ if $\boldsymbol{\Psi}$ approaches zero so that $(\boldsymbol{\Lambda}'\boldsymbol{\Psi}^{-2}\boldsymbol{\Lambda})^{-1}$ approaches zero. It is less obvious that if one element per column in $\boldsymbol{\Lambda}$ approaches one so that the corresponding element in $\boldsymbol{\Psi}$ approaches zero, the diagonal elements in $\boldsymbol{\Lambda}'\boldsymbol{\Psi}^{-2}\boldsymbol{\Lambda}$ will approach infinity so that $(\boldsymbol{\Lambda}'\boldsymbol{\Psi}^{-2}\boldsymbol{\Lambda})^{-1}$ approaches zero, which implies that Equation (7) approaches $\boldsymbol{\Lambda}\boldsymbol{\Phi}\boldsymbol{\Lambda}'$. This implies that conventional factor score predictors reproduce the common part of the original covariance matrix exactly, when there is one variable per factor with a zero error variance.

However, it will rarely be the case that a variable has a zero error variance. Therefore, the aim of the present study is to investigate whether a single variable factor score predictor with a nonzero error variance can –under some conditions– reproduce the common part of the observed covariances more exactly than conventional factor score predictors. It should be noted that especially the non-diagonal elements of the reproduced covariances are in the focus of interest. The reason is that the non-diagonal elements represent the covariances between different variables, which are the basis for the concept of common variance. Under this focus the aim is to find values for $\mathbf{B}$, for which

$$\mathbf{\Lambda\Phi\Lambda'} - \text{diag}(\mathbf{\Lambda\Phi\Lambda'}) = \mathbf{\Sigma B(B'\Sigma B)^{-1}B'\Sigma} - \text{diag}(\mathbf{\Sigma B(B'\Sigma B)^{-1}B'\Sigma}). \qquad (8)$$

If the equality of the non-diagonal elements of the reproduced covariance matrices cannot be achieved the mean squared difference between the reproduced covariances (see Theorem 3.2 and Equation 20) might be minimized. For the unrealistic case that there are as many factors as variables and that each factor is represented by a single variable we have $\mathbf{B} = \mathbf{I}$, which implies that Equation 8 can be written as

$$\mathbf{\Lambda\Phi\Lambda'} - \text{diag}(\mathbf{\Lambda\Phi\Lambda'}) = \mathbf{\Sigma} - \text{diag}(\mathbf{\Sigma}), \qquad (9)$$

which is true because

$$\mathbf{\Lambda\Phi\Lambda'} - \text{diag}(\mathbf{\Lambda\Phi\Lambda'}) = \mathbf{\Lambda\Phi\Lambda'} + \mathbf{\Psi^2} - \text{diag}(\mathbf{\Lambda\Phi\Lambda'} + \mathbf{\Psi^2}). \qquad (10)$$

In order to inspect a more realistic condition, consider a perfect (block diagonal) orthogonal simple structure with two variables with nonzero loadings on each factor ($p/q = 2$) and zero loadings on all other factors (the spherical case). Accordingly, $\mathbf{\Lambda}$ can be decomposed into $q$ submatrices $\mathbf{\Lambda_i}$ with 2 x 1 nonzero elements, so that the block diagonal loading pattern can be written as

$$\mathbf{\Lambda} = \begin{bmatrix} \mathbf{\Lambda_i} & \cdots & 0 \\ \vdots & \ddots & \vdots \\ 0 & \cdots & \mathbf{\Lambda_q} \end{bmatrix}. \qquad (11)$$

On this basis the decomposition of observed covariances can be written as

$$\mathbf{\Sigma} = \begin{bmatrix} \mathbf{\Sigma_i} & \cdots & 0 \\ \vdots & \ddots & \vdots \\ 0 & \cdots & \mathbf{\Sigma_q} \end{bmatrix} = \begin{bmatrix} \mathbf{\Lambda_i\Lambda_i'} & \cdots & 0 \\ \vdots & \ddots & \vdots \\ 0 & \cdots & \mathbf{\Lambda_q\Lambda_q'} \end{bmatrix} + \begin{bmatrix} \mathbf{\Psi_i^2} & \cdots & 0 \\ \vdots & \ddots & \vdots \\ 0 & \cdots & \mathbf{\Psi_q^2} \end{bmatrix}, \text{ with } i = 1 \text{ to } q. \qquad (12)$$

In Equation 12 $\mathbf{\Sigma_i}$ to $\mathbf{\Sigma_q}$ are the 2 x 2 submatrices of observed covariances and $\mathbf{\Psi_i^2}$ to $\mathbf{\Psi_q^2}$ are the 2 x 2 submatrices of covariances for the error scores. On this basis it is possible to show that taking one of the variables in each submatrix $\mathbf{\Lambda_i}$ as the score predictor implies that the condition expressed in Equation (8) also holds if $\mathbf{\Sigma}$ is a correlation matrix (Theorem 3.1).

**Theorem 3.1.** $\mathbf{\Lambda\Lambda'} - \text{diag}(\mathbf{\Lambda\Lambda'}) = \mathbf{\Sigma B(B'\Sigma B)^{-1}B'\Sigma} - \text{diag}(\mathbf{\Sigma B(B'\Sigma B)^{-1}B'\Sigma})$ *if $p/q = 2$,*

$\text{diag}(\mathbf{\Sigma}) = \mathbf{I}, \mathbf{\Lambda} = \begin{bmatrix} \mathbf{\Lambda_i} & \cdots & 0 \\ \vdots & \ddots & \vdots \\ 0 & \cdots & \mathbf{\Lambda_q} \end{bmatrix}, \mathbf{B}_i = \begin{bmatrix} 1 \\ 0 \end{bmatrix}$, *and* $\mathbf{B} = \mathbf{I}_q \otimes \mathbf{B}_i$,

where $\mathbf{I}_q$ is a $q$ x $q$ identity matrix and "$\otimes$" denotes the Kronecker-product.

*Proof.* The submatrix $\mathbf{\Sigma_i}$ can be written as $\mathbf{\Sigma}_i = \begin{bmatrix} 1 & \sigma_{12} \\ \sigma_{21} & 1 \end{bmatrix}$ with $\sigma_{21} = \sigma_{12}$. For the submatrices

Equation 8 can be written as

$$\mathbf{\Lambda}_i\mathbf{\Lambda}_i{}' - \text{diag}(\mathbf{\Lambda}_i\mathbf{\Lambda}_i{}') = \mathbf{\Sigma}_i\mathbf{B}_i(\mathbf{B}_i{}'\mathbf{\Sigma}_i\mathbf{B}_i)^{-1}\mathbf{B}_i{}'\mathbf{\Sigma}_i - \text{diag}(\mathbf{\Sigma}_i\mathbf{B}_i(\mathbf{B}_i{}'\mathbf{\Sigma}_i\mathbf{B}_i)^{-1}\mathbf{B}_i{}'\mathbf{\Sigma}_i). \tag{13}$$

It follows from $\mathbf{B}_i = \begin{bmatrix} 1 \\ 0 \end{bmatrix}$ and $\text{diag}(\mathbf{\Sigma}) = \mathbf{I}$ that $(\mathbf{B}_i{}'\mathbf{\Sigma}_i\mathbf{B}_i)^{-1} = \mathbf{1}$ so that $\mathbf{B}_i(\mathbf{B}_i{}'\mathbf{\Sigma}_i\mathbf{B}_i)^{-1}\mathbf{B}_i{}' = \begin{bmatrix} 1 & 0 \\ 0 & 0 \end{bmatrix}$ and

$\mathbf{\Sigma}_i\mathbf{B}_i(\mathbf{B}_i{}'\mathbf{\Sigma}_i\mathbf{B}_i)^{-1}\mathbf{B}_i{}' = \begin{bmatrix} 1 & 0 \\ \sigma_{21} & 0 \end{bmatrix}$. This implies

$$\mathbf{\Sigma}_i\mathbf{B}_i(\mathbf{B}_i{}'\mathbf{\Sigma}_i\mathbf{B}_i)^{-1}\mathbf{B}_i{}'\mathbf{\Sigma}_i = \begin{bmatrix} 1 & 0 \\ \sigma_{21} & 0 \end{bmatrix}\begin{bmatrix} 1 & \sigma_{12} \\ \sigma_{21} & 1 \end{bmatrix} = \begin{bmatrix} 1 & \sigma_{12} \\ \sigma_{21} & \sigma_{21}\sigma_{12} \end{bmatrix} \tag{14}$$

and, accordingly, that Equation (13) is true since the non-diagonal elements of $\mathbf{\Sigma}_i$ and $\mathbf{\Sigma}_i\mathbf{B}_i(\mathbf{B}_i{}'\mathbf{\Sigma}_i\mathbf{B}_i)^{-1}\mathbf{B}_i{}'\mathbf{\Sigma}_i$ are equal. This completes the proof because this can be shown for any submatrix $\mathbf{\Lambda}_i$ to $\mathbf{\Lambda}_q$ and $\mathbf{\Sigma}_i$ to $\mathbf{\Sigma}_q$. ☐

Accordingly, for orthogonal models with a block-diagonal loading matrix based on zero-loadings and three variables with non-zero or salient loadings on each factor the decomposition of $\mathbf{\Sigma}$ can also be based on Equation (12). The selection of a single variable yields $\mathbf{B}_i{}' = \begin{bmatrix} 1 & 0 & 0 \end{bmatrix}$ so that

$$\mathbf{B}_i(\mathbf{B}_i{}'\mathbf{\Sigma}_i\mathbf{B}_i)^{-1}\mathbf{B}_i{}' = \begin{bmatrix} 1 & 0 & 0 \\ 0 & 0 & 0 \\ 0 & 0 & 0 \end{bmatrix}, \ \mathbf{\Sigma}_i = \begin{bmatrix} 1 & \sigma_{12} & \sigma_{13} \\ \sigma_{21} & 1 & \sigma_{23} \\ \sigma_{31} & \sigma_{32} & 1 \end{bmatrix}, \text{ and}$$

$$\mathbf{\Sigma}_i\mathbf{B}_i(\mathbf{B}_i{}'\mathbf{\Sigma}_i\mathbf{B}_i)^{-1}\mathbf{B}_i{}'\mathbf{\Sigma}_i = \begin{bmatrix} 1 & \sigma_{12} & \sigma_{13} \\ \sigma_{21} & \sigma_{21}\sigma_{12} & \sigma_{21}\sigma_{13} \\ \sigma_{31} & \sigma_{31}\sigma_{12} & \sigma_{31}\sigma_{13} \end{bmatrix}, \tag{15}$$

which implies that four non-diagonal elements of $\mathbf{\Sigma}_i$ are reproduced exactly and that two non-diagonal elements of $\mathbf{\Sigma}_i$ are not reproduced exactly, unless $\sigma_{32} = \sigma_{31}\sigma_{12}$. The latter is possible but very unlikely, so that the proportion of exactly reproduced elements that occurs in each submatrix can be calculated from the number of non-zero or salient loadings per factor, which is given by $p/q$ in a model with perfect simple structure. Thus, for block-diagonal orthogonal loading patterns, the number of exactly reproduced non-diagonal elements is proportional to the number of non-diagonal elements that are in the row and in the column of the variable that is used as factor score predictor, so that it is proportional to $2(p/q - 1)$. In contrast, the number of not exactly reproduced elements is proportional to the remaining non-diagonal elements, so that it is proportional to $(p/q - 1)(p/q - 2)$. This implies that the number of not exactly reproduced non-diagonal elements increases $0.5(p/q - 2)$ times faster with increasing number of salient loadings than the number of exactly reproduced elements.

Although it is impossible to give an account of the relative precision of reproducing the non-diagonal correlations for the single variable factor score predictor and conventional factor score predictors by means of Equation (15), it is possible to prove that the single variable factor score predictor reproduces the non-diagonal correlations more precisely than the conventional factor score predictor for orthogonal perfect simple structures with $p/q = 3$ under the condition of constant salient loadings below a certain size $h$ on each factor, which can be expressed as $\mathbf{\Lambda}_i = \sigma^{1/2}\mathbf{1}$ with $\sigma^{1/2} < h$, where $\mathbf{1}$ is a unit-vector.

**Theorem 3.2.** $SSQ(\mathbf{\Sigma} - \mathbf{\Sigma B}(\mathbf{B}'\mathbf{\Sigma B})^{-1}\mathbf{B}'\mathbf{\Sigma} - \text{diag}(\mathbf{\Sigma} - \mathbf{\Sigma B}(\mathbf{B}'\mathbf{\Sigma B})^{-1}\mathbf{B}'\mathbf{\Sigma})) < SSQ(\mathbf{\Sigma} - \mathbf{\Sigma_r} - \text{diag}(\mathbf{\Sigma} - \mathbf{\Sigma_r}))$, *if* $p/q = 3$ *and* $\mathbf{\Lambda}_i = \sigma^{1/2}\mathbf{1}$ *with* $\sigma^{1/2} < h$, $\text{diag}(\mathbf{\Sigma}) = \mathbf{I}$, $\mathbf{\Lambda} = \begin{bmatrix} \mathbf{\Lambda}_i & \cdots & 0 \\ \vdots & \ddots & \vdots \\ 0 & \cdots & \mathbf{\Lambda}_q \end{bmatrix}$, $\mathbf{B}_i = \begin{bmatrix} 1 \\ 0 \\ 0 \end{bmatrix}$, *and* $\mathbf{B} = \mathbf{I}_q \otimes \mathbf{B}_i$, *with* $SSQ$ *denoting the sum of squares.*

*Proof.* It follows from Equation (15) and from $\mathbf{\Lambda}_i = \sigma^{1/2}\mathbf{1}$ that

$$SSQ(\mathbf{\Sigma}_i - \mathbf{\Sigma}_i\mathbf{B}_i(\mathbf{B}_i'\mathbf{\Sigma}_i\mathbf{B}_i)^{-1}\mathbf{B}_i'\mathbf{\Sigma}_i - \text{diag}(\mathbf{\Sigma}_i - \mathbf{\Sigma}_i\mathbf{B}_i(\mathbf{B}_i'\mathbf{\Sigma}_i\mathbf{B}_i)^{-1}\mathbf{B}_i'\mathbf{\Sigma}_i)) = 2(\sigma - \sigma^2)^2 \qquad (16)$$

for each block of non-zero correlations.

It follows from Equation (7) and from $\mathbf{\Lambda}_i = \sigma^{1/2}\mathbf{1}$ that

$$\mathbf{\Lambda}_i((\mathbf{\Lambda}_i'\mathbf{\Psi}_i^{-2}\mathbf{\Lambda}_i)^{-1} + 1)\mathbf{\Lambda}_i' = \sigma^{1/2}\mathbf{1}((\sigma^{1/2}\mathbf{1}'\mathbf{\Psi}_i^{-2}\mathbf{1}\sigma^{1/2})^{-1} + 1)\mathbf{1}'\sigma^{1/2}$$

$$= \mathbf{1}(\sigma^{1/2}(\sigma^{1/2}\mathbf{1}'\mathbf{\Psi}_i^{-2}\mathbf{1}\sigma^{1/2})^{-1}\sigma^{1/2} + \sigma)\mathbf{1}' = \mathbf{1}((\mathbf{1}'\mathbf{\Psi}_i^{-2}\mathbf{1})^{-1} + \sigma)\mathbf{1}'$$

$$= \mathbf{1}((3(1 - \sigma)^{-1})^{-1} + \sigma)\mathbf{1}' = \mathbf{1}(1/3 - 1/3\sigma + \sigma)\mathbf{1}'. \qquad (17)$$

Equation (17) yields the non-diagonal elements of $\mathbf{\Sigma}_{ir}$. Since $\sigma$ represents the non-diagonal elements of $\mathbf{\Sigma}_i$ and since there are six non-diagonal elements in $\mathbf{\Sigma}_i$ Equation (17) implies that

$$SSQ(\mathbf{\Sigma}_i - \mathbf{\Sigma}_{ir} - \text{diag}(\mathbf{\Sigma}_i - \mathbf{\Sigma}_{ir})) = 6(1/3 - 1/3\sigma)^2. \qquad (18)$$

Equating (16) and (18) and some transformation yields

$$\sigma^2(2^{1/2} - 2^{1/2}\sigma)^2 = 1/3(2^{1/2} - 2^{1/2}\sigma)^2, \qquad (19)$$

which implies that the left side is smaller than the right side of Equation (19) for $\sigma^2 < 1/3$, that is, $\sigma^{1/2} < 1/3^{1/4} = 0.76 = h$. This completes the proof. $\square$

Thus, for orthogonal population models based on three equal loadings per factor smaller than 0.76 the correlations are more precisely reproduced from the single variable factor score predictor than from conventional factor score predictors. Such thresholds for loadings or correlations could be found for different numbers of salient loadings per factor, but it is more efficient to identify these conditions by means of the simulation of population models (see below).

It should be noted that the superiority of the single variable factor score predictor is not so much based on the high precision of the reproduction of non-diagonal covariances based on this score, but on the low precision of conventional factor score predictors in reproducing the non-diagonal elements of the covariance matrix, especially, when the absolute size and the number of non-diagonal elements are small. This can be shown for the one-factor model based on variables with equal loadings (Theorem 3.3), for which the loading matrix can be written as $\boldsymbol{\Lambda} = l\mathbf{1}$, where $l$ is a scalar and $\mathbf{1}$ a $p$ x 1 unit-vector.

**Theorem 3.3.** $\lim_{l \to 0} (\boldsymbol{\Lambda}) = \mathbf{0}$ $implies$ $(\boldsymbol{\Sigma} - diag(\boldsymbol{\Sigma})) = \mathbf{0}$ $and$ $\boldsymbol{\Lambda}(\boldsymbol{\Lambda}'\boldsymbol{\Sigma}^{-1}\boldsymbol{\Lambda})^{-1}\boldsymbol{\Lambda}' = p^{-1}\mathbf{1}\mathbf{1}'$,

$for$ $diag(\boldsymbol{\Sigma}) = \mathbf{I}$.

*Proof.* It follows from Equation (2) and from $diag(\boldsymbol{\Sigma}) = \mathbf{I}$ that $\lim_{l \to 0} (\boldsymbol{\Lambda}) = \mathbf{0}$ implies $\boldsymbol{\Sigma} = \mathbf{I}$ and, accordingly,

$$\boldsymbol{\Lambda}(\boldsymbol{\Lambda}'\boldsymbol{\Sigma}^{-1}\boldsymbol{\Lambda})^{-1}\boldsymbol{\Lambda}' = l\mathbf{1}(l\mathbf{1}'\boldsymbol{\Sigma}^{-1}l\mathbf{1})^{-1}l\mathbf{1}'l = \mathbf{1}(\mathbf{1}'\mathbf{1})^{-1}\mathbf{1}' = p^{-1}\mathbf{1}\mathbf{1}'. \tag{20}$$

This completes the proof. □

Thus, whereas the non-diagonal elements of $\boldsymbol{\Sigma}$ are zero in the limit, the non-diagonal elements of $\boldsymbol{\Lambda}(\boldsymbol{\Lambda}'\boldsymbol{\Sigma}^{-1}\boldsymbol{\Lambda})^{-1}\boldsymbol{\Lambda}'$ are $p^{-1}$. The fact that the non-diagonal elements are zero implies that taking a single variable as a factor score predictor will correctly reproduce the matrix of (non-diagonal) zero inter-correlations. The extreme case of zero inter-correlations is, of course, meaningless. However, it follows from Theorem 3.3 that conventional factor score predictors overestimate the size of the non-diagonal elements when the loadings an $p$ are small.

Another implication of Equation (15) is that the choice of the non-diagonal elements that can be exactly reproduced could be performed in a way that minimizes $|\sigma_{32} - \sigma_{31}\sigma_{12}|$. Choosing the variable with the largest absolute loading implies that the largest correlations are exactly reproduced and that $|\sigma_{32} - \sigma_{31}\sigma_{12}|$ is a minimum. In orthogonal solutions with perfect simple structure this implies that for each factor the variable with the smallest uniqueness is chosen as a factor score predictor. However, sampling error affects the size of the common loadings and unique loadings in the sample so that it is not clear to what degree the abovementioned results also hold in the sample. In order to investigate the abovementioned relations between *p/q* and the precision of non-diagonal covariance reproduction a simulation study for a large number of different population models as well as for samples drawn from a subset of population models was performed.

## 4 Simulation Studies

### 4.1 Simulation for populations

Simulations that are restricted to population models (without sampling error) can be used in order to determine the effects of specific conditions or predictions that might follow from algebraic considerations (e.g. Widaman, 2007). The simulation studies were performed with IBM SPSS Version 21. The present population simulation study was based on four sets of typical population models: Model set 1 comprised orthogonal population models with equal size of salient loadings (*l*) and zero non-salient loadings. Model set 2 comprised orthogonal population models with variable size of salient loadings and zero non-salient loadings. In these models, each half of the loadings were .10 above or below the mean salient loading so that the maximal difference between the salient loadings was .20. Model set 3 comprised oblique population models with equal size of salient loadings and zero non-salient loadings and inter-factor correlations of .40. Model set 4 comprised oblique population models with unequal size of salient loadings (exactly like in Model set 2), zero non-salient loadings and inter-factor correlations of .40. The number of factors (*q*) varied between 1 and 10 and the number of salient loadings per factor (*p/q*) varied between 2 and 10. The mean of the salient loadings (*l*) varied between .25 and .95 in steps of .05 for Model sets 1 and 3, it varied between .25 and .85 for Model sets 2 and 4. The variation of the mean salient loadings in Model sets 2 and 4 was smaller in order to avoid loadings greater one.

The simulation study was based on the following indicators: The mean squared difference for the non-diagonal elements of the matrix of observed covariances and the non-diagonal elements of the covariances reproduced from conventional factor score predictors is

$$\Delta_{\mathbf{r}} = MSQ((\mathbf{\Sigma} - \mathbf{\Sigma_r}) - \mathrm{diag}(\mathbf{\Sigma} - \mathbf{\Sigma_r})), \tag{21}$$

where *MSQ* is the mean of all squared matrix elements. Thus, $\Delta_{\mathbf{r}}$ represents the mean squared error of non-diagonal covariance reproduction for conventional factor score predictors. The mean squared difference for the non-diagonal elements of the matrix of observed covariances and the non-diagonal elements of the covariances reproduced from the single variable score is

$$\Delta_{\mathbf{b}} = MSQ((\Sigma - \Sigma_{\mathbf{b}}) - \text{diag}(\Sigma - \Sigma_{\mathbf{b}})), \tag{22}$$

so that $\Delta_{\mathbf{b}}$ represents the mean squared error of non-diagonal covariance reproduction for single variable scores. The difference $\Delta_{\mathbf{r}} - \Delta_{\mathbf{b}}$ is positive when $\Delta_{\mathbf{r}} > \Delta_{\mathbf{b}}$ and it is negative when $\Delta_{\mathbf{b}} > \Delta_{\mathbf{r}}$. Thus, negative values indicate that conventional factor score predictors give a more precise account (have less error) of the non-diagonal elements of the observed covariance matrix than the single variable score.

The results for Model set 1 indicate that conventional factor score predictors have less error of non-diagonal covariance reproduction than single variable scores only when salient loadings ($l$) are large and when the number of salient loadings per factor ($p/q$) is large (see Figure 1 A). The pattern of results was extremely similar for Model set 3 so that an additional Figure on the relation between $l$ and $p/q$ for these models was not necessary. However, for Model set 2 the overall error of non-diagonal covariance reproduction was slightly reduced, both for conventional factor score predictors and single variable scores (see Figure 1 B). Again, the pattern of results was extremely similar for Model sets 2 and 4 so that a Figure for Model set 4 was not necessary. There was no effect of the number of factors ($q$) on the error of non-diagonal covariance reproduction so that the results in Figure 1 were collapsed across $q$.

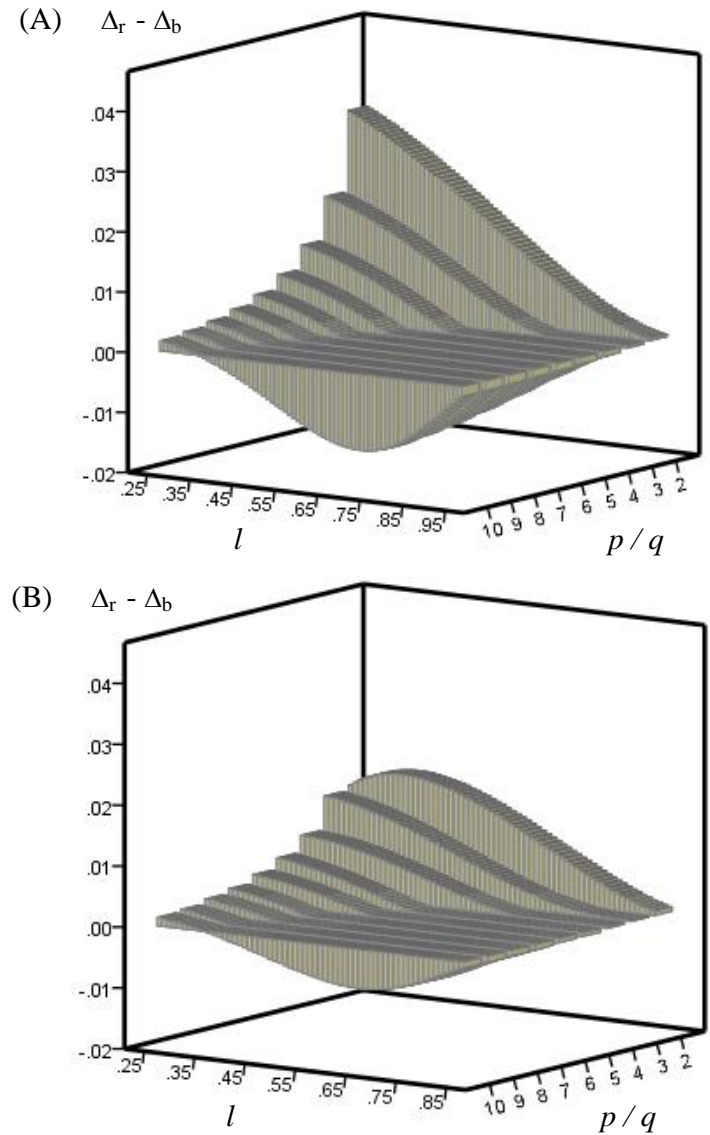

Figure 1.

Difference between the mean squared error of non-diagonal covariance reproduction for conventional factor score predictors ($\Delta_r$) and the mean squared error of non-diagonal covariance reproduction for single variable scores ($\Delta_b$) for population Model set 1 (A) and population Model set 3 (B).

However, in order to compare the effects for the different model sets, the salient loading size and the number of salient loadings per factor, below which the single variable score has a more precise non-diagonal covariance reproduction than the conventional factor score predictors is presented in Figure 2. It should be noted that the maximum loading of .95 for Model sets 1 and 3 and a maximum loading of .85 for Model sets 2 and 4 occurred because this was the maximum of the salient loadings that were used for the population models.

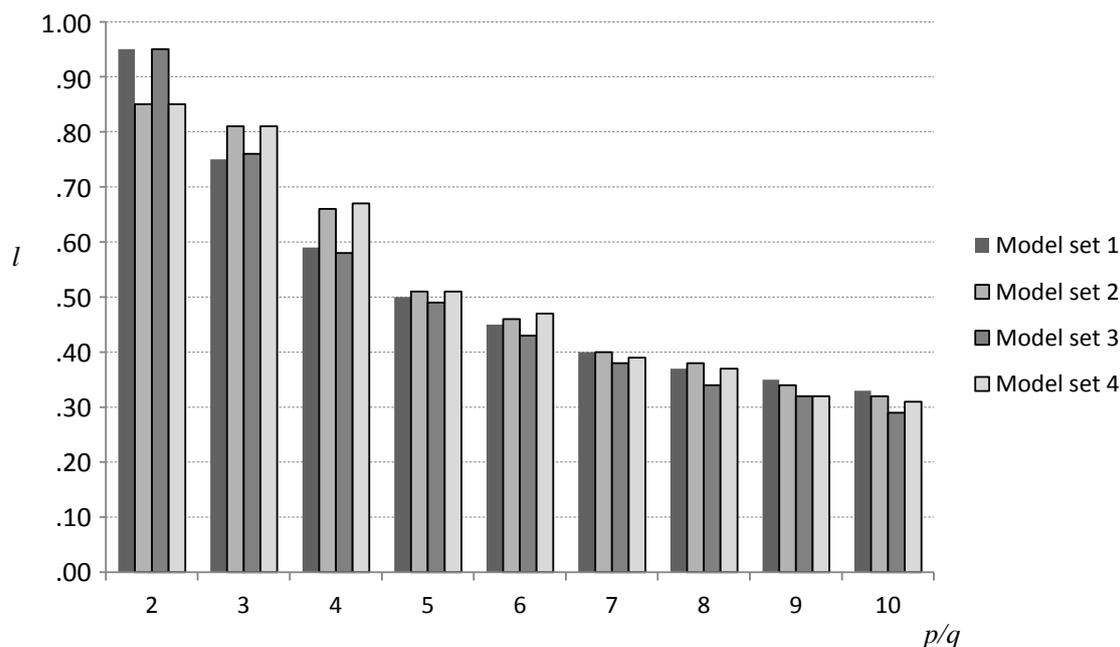

Figure 2.
Salient loading size below which single variable scores had less mean squared error in non-diagonal covariance reproduction than conventional factor score predictors.

Interestingly, the differences between the model sets are rather small. Although the conceptual differences between the model sets were not extreme (uncorrelated factors versus inter-factor correlation of .40 and constant salient loadings versus maximal differences of .20 between the salient loadings), this indicates that for several models the precision of non-diagonal covariance reproduction can be anticipated from the number of salient loadings per factor ($p/q$) and from the mean salient loading size ($l$). As a rule of thumb one can conclude that having small (absolute) salient loadings ($l \leq .40$) and six or less salient loadings per factor ($p/q \leq 6$) indicates that single variable scores yield a more precise reproduction of non-diagonal covariances than conventional factor score predictors. In contrast, when salient loadings are larger ($l > .60$) and when there are many salient variables per factor ($p/q > 6$) conventional factor score predictors yield a more precise reproduction of non-diagonal covariances than single variable scores.

## 4.2  Simulation for samples

In order to investigate the effect of sampling error on the results obtained for the population models a subset of the models presented in the simulation for the population was analyzed for samples. Six variables were manipulated in order to represent the conditions of the study: The

(1) number of factors ($q$) was 1, 3 or 9, the (2) number of variables per factor ($p/q$) was 3 to 10, three levels of (3) mean salient population loadings were selected ($l$ = .40, .60, .80) whereas non-salient loadings were zero. (4) For all factors of each simulation condition the salient loadings were either constant or variable, where each half of the loadings were .10 above or below the mean salient loading. (5) The inter-factor correlations were zero or .40 and three levels of (6) sample sizes were chosen ($n$ = 150, 300, 900). For each of the 720 conditions (5 models with 1, 3 or 9 orthogonal versus oblique factors x 8 number of variables per factor x 3 loading size x 2 constant versus variable loadings x 2 orthogonal versus oblique models x 3 sample sizes) 5,000 data sets were generated by means of random variables from a multivariate normal distribution and submitted to unweighted-least squares (ULS) factor analysis. Varimax-rotation was performed for the orthogonal models and Promax-rotation was performed for the oblique models.

The difference between the precision of non-diagonal covariance reproduction for single variables and conventional factor score predictors in orthogonal models is presented in Figure 3. For the orthogonal models with constant salient loadings (Figure 3A) and variable salient loadings (Figure 3B), single variable scores resulted in more precise non-diagonal covariance reproduction than the conventional factor score predictors when the mean salient loadings were small ($l$ = .40), when $p/q \leq 6$ and the number of factors ($q$) was small (1-3). This is the rule of thumb that has already been obtained for the population data. Accordingly, the effects corresponding to this rule were more pronounced for larger sample sizes than for smaller sample sizes. An unexpected but notable result was that the difference of the mean squared error of non-diagonal covariance reproduction approaches zero for 9 factors (see Figure 3). This indicates that for a large number of factors the precision of reproduction of non-diagonal elements is nearly identical for conventional factor score predictors and single variables. The results were similar for the simulations based in population models with inter-factor correlations of .40 so that no additional figure was necessary.

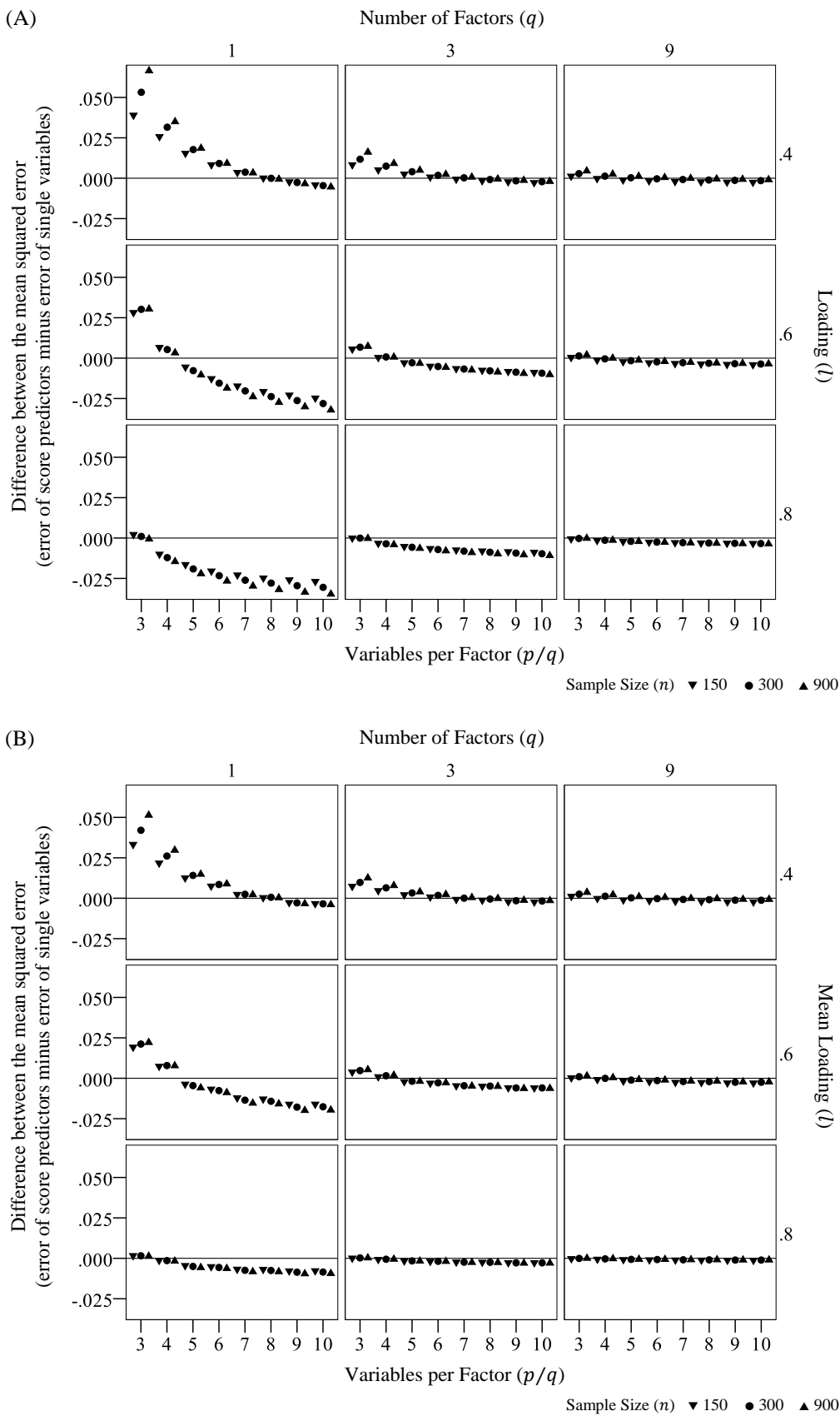

Figure 3.
Sample based simulation of the mean difference between the mean squared error of non-diagonal covariance reproduction for conventional factor score predictors ($\Delta_r$) and single variable scores ($\Delta_b$) for orthogonal models based on constant salient loadings (A) and variable salient loadings (B).

## 5 Discussion

The paper starts from the fact that –even when the common factor model holds– the observed covariances reproduced from the regression components corresponding to conventional factor score predictors like Thurstone's, Bartlett's, Anderson-Rubin's, and McDonald's factor score predictor is not identical to the covariances reproduced from the common factors (Beaducel, 2007). This result is specified further by showing that this result holds, unless the uniqueness of at least one variable with a salient loading on each factor is zero. This specification corresponds to Spearman's (1929) idea to use a single variable with perfect communality as a factor score predictor. However, variables with perfect communality are rare so that factor score predictors will generally not precisely reproduce the non-diagonal observed covariances. This may be regarded as a problem of conventional factor score predictors, because the common factors themselves perfectly reproduce the common part of the observed covariances when the common factor model holds. Therefore, the similarity of the common variances of observed variables reproduced from the factors and reproduced from the factor score predictors can be regarded as an additional criterion for the evaluation of factor score predictors. This criterion considers whether the factor score predictors can be used as substitutes for the factors in order to explain the common variances and was therefore named structural similarity. This justification is, of course, more theoretical, but, whenever factors are interpreted as being the cause of covariances, this criterion would be appropriate. The multiple correlation between the variables and the factor (validity) does not help here, because the roles are inverted in this criterion since the observed variables are used as predictors of the factor, so that the factor score predictor is not investigated as being a cause of the observed covariances.

It was, moreover, shown that under some conditions the difference between the covariances reproduced from the common factors and the covariances reproduced from factor score predictors can be minimized by means of choosing a single variable with a minimal uniqueness. For example, it was shown for orthogonal factor models with a perfect (block-diagonal) simple structure based on two variables per factor that scores based on choosing a single variable per factor lead to a perfect reproduction of the non-diagonal covariances. It was also shown that for a simple structure with three constant salient (non-zero) loadings below a certain value a single variable factor score predictor reproduces the non-diagonal correlations more precisely than the conventional factor score predictors. Moreover, it was shown that the precision of the reproduction of the non-diagonal covariances decreases with the number of variables per factor. Finally, it was shown that the larger precision of the single variable factor score predictor compared to conventional factor score predictors is primarily due to the very low precision of the reproduction of non-diagonal correlations from conventional factor score predictors when the observed correlations are small. From a more general point of view it should be noted that the focus on the non-diagonal elements of the covariance matrix has already been proposed in the context of the Minres-method of factor analysis (Harman, 1976). Thus, the present investigation of single-variable scores was performed for the Minres-criterion of minimizing the non-diagonal residuals of the covariance matrix.

The attempt of identifying a single variable with maximal structural similarity to the factor score can be regarded as a relaxed version of Spearman's idea to use a single variable with perfect communality. However, the aim of Spearman (1929) was to eliminate indeterminacy of factor scores whereas here, the less ambitious aim was to reproduce the non-diagonal observed covariances with maximal precision. It should be noted that the criterion of structural similarity of factor score predictors and factor scores does not solve or circumvent the problem of factor score indeterminacy.

Accordingly, a population simulation study based on several different population models showed that the precision of non-diagonal covariance reproduction can be anticipated from the number of salient loadings per factor and from the mean salient loading size. As a rule of thumb one can conclude that having small salient loadings ($\leq .40$) and six or less salient loadings per factor indicates that single variable scores yield a more precise reproduction of non-diagonal covariances than conventional factor score predictors. This rule of thumb was also found in a simulation study based on small (N=150), medium (N=300), and large (N=900) sample sizes and ULS factor analysis. The fact that using a single variable as factor score predictor does not lead to optimal structural similarity under all conditions investigated in the simulation study implies that other factor score predictors, probably not based on single variables, might be developed in order to achieve optimal structural similarity for those conditions where the single variable approach fails. Nevertheless, the condition of a small number of variables with rather small salient loadings might regularly occur in applied research so that using a single variable with highest communality as a factor score predictor could be regarded as an interesting possibility under several conditions. It should be noted that other simplified procedures for the computation of factor score predictors have already been shown to work quite well (Grice, 2001b), which underlines that short-cuts of the complexities of the computation of factor score predictors are possible.

Moreover, it was found in the simulation based on samples that a single variable has nearly the same precision of non-diagonal covariance reproduction than conventional factor score predictors when the number of common factors is large. Thus, with nine factors no relevant difference in the reproduction of non-diagonal covariances occurred between single variable scores and the conventional factor score predictors. This indicates that under these conditions the conventional factor score predictors should be preferred, because they are optimal according to other criteria (e.g. validity).

Overall, the results can be regarded as an extension of the discussion that emerged between Spearman (1929) and Wilson (1929), when Spearman (1929) proposed to add a single variable with a perfect loading on the general intelligence factor in order to avoid indeterminacy of the factor score for general intelligence. Wilson (1929) responded that in this cases the single variable with the perfect loading will already represent the factor score of general intelligence. The present result adds to this discussion that –for a small number of variables per factor and small salient loadings– choosing a single variable with a maximal loading may represent a good factor score predictor in the sense that the non-diagonal covariances reproduced from this score will be more similar to the observed non-diagonal covariances than the covariances reproduced from conventional factor score predictors. From this perspective, common factor analysis could be used as a method that allows to identify those variables for which the reproduced non-diagonal covariances are most similar to the observed non-diagonal covariances.

As mentioned in the introduction, it should be acknowledged that besides the reproduction of the non-diagonal elements of the covariances many other aspects of factor score predictors have previously been emphasized (Mulaik, 2010; Beauducel & Rabe, 2009; Krijnen, 2006; Grice, 2001a; McDonald & Burr, 1967). Especially, the correlation of the factor score predictor with the factor score might be regarded as more relevant than the optimal reproduction of the non-diagonal elements of the covariance matrix. However, if one considers that the focus of common factor loadings, especially in contrast to principal component loadings, is on the reproduction of the non-diagonal elements of the observed covariances, one might ask whether the correlation of the factor score predictor with the factor scores is really as important, when it is possible to show that under some conditions, single variable scores reproduce the observed covariances even better than conventional factor score predictors that are based on the correlation between the factor score and the factor score predictor.